\documentclass[a4paper,12pt]{article}
\usepackage{graphicx}
\usepackage[footnotesize]{caption}

\usepackage{multirow}
\usepackage{bm}
\usepackage{bbm}
\usepackage{slashed}
\usepackage{mathtools}

\usepackage{axodraw}

\newcommand{\rr}[1]{\mathrm{#1}}

\def\beq{\begin{equation}}
\def\eeq{\end{equation}}
\def\bea{\begin{eqnarray}}
\def\eea{\end{eqnarray}}

\def\<{\left\langle}
\def\>{\right\rangle}

\newcommand{\bc}{\begin{center}}
\newcommand{\ec}{\end{center}}
\newcommand{\bd}{\begin{displaymath}}
\newcommand{\ed}{\end{displaymath}}
\newcommand{\be}{\begin{equation}}
\newcommand{\ee}{\end{equation}}
\newcommand{\ba}{\begin{array}}
\newcommand{\ea}{\end{array}}
\newcommand{\bt}{\begin{tabular}}
\newcommand{\et}{\end{tabular}}

\newcommand{\ds}{\displaystyle}

\begin{document}

\bibliographystyle{OurBibTeX}

\begin{titlepage}

\begin{center}
{ \sffamily \Large Neutralino Dark Matter with Inert Higgsinos and Singlinos}
\\[8mm]
Jonathan~P.~Hall and Stephen~F.~King
\\[3mm]
{\small\it
School of Physics and Astronomy, University of Southampton,\\
Southampton, SO17 1BJ, U.K.}
\\[1mm]
\end{center}
\vspace*{0.75cm}

\begin{abstract}

\noindent
We discuss neutralino dark matter arising from supersymmetric models with extra inert Higgsinos
and singlinos, where inert means that their scalar partners do not get vacuum expectation values.
As an example, we consider the
extended neutralino sector of the E$_6$SSM, which predicts three
families of Higgs doublet pairs, plus three singlets, plus a $Z'$, together
with their fermionic superpartners.
We show that the two families of inert doublet Higgsinos
and singlinos predicted by this model provide an almost
decoupled neutralino sector with a naturally light LSP which can account for
the cold dark matter relic abundance independently of the
rest of the model, providing that the ratio of the two usual
Higgs doublets satisfies $\tan \beta < 2$.
\end{abstract}

\end{titlepage}
\newpage
\setcounter{footnote}{0}

\section{Introduction}

The existence of weak scale supersymmetry (SUSY) is theoretically
well motivated, because of its ability to stabilise the electro-weak
symmetry breaking (EWSB) scale.
One of the benefits of weak scale SUSY with conserved $R$-parity
is that the lightest supersymmetric particle (LSP) is absolutely
stable and provides a weakly interacting massive particle (WIMP)
candidate capable of accounting for
the observed cold dark matter (CDM) relic density
$\Omega_\rr{CDM} h^2\approx 0.1$ \cite{EHNOS,hep-ph/9506380}.
In particular, the lightest neutralino in SUSY models
is an excellent such candidate, providing its mass,
composition and interactions are suitably tuned to result in
the correct value of $\Omega_\rr{CDM} h^2$.

The Minimal Supersymmetric Standard Model (MSSM)
\cite{Chung:2003fi} provides the simplest supersymmetric
extension of the Standard Model (SM) in which the superpotential
contains the bilinear term $\mu H_{d}H_{u}$, where ${H}_{d,u}$ are
the two Higgs doublets whose neutral components $H_{d,u}^0$
develop vacuum expectation values (VEVs) at the weak scale and the
$\mu$ parameter has the dimensions of mass. However, since this term respects
supersymmetry, there is no reason for $\mu$ to be of order the weak
scale, leading to the so-called $\mu$ problem \cite{Cohen:2008ni}.
Also, the MSSM suffers a fine-tuning of parameters at the
per cent level \cite{Kane:1998im}.

To address the above shortcomings of the
MSSM one may replace the $\mu$ term of the MSSM
by the low energy VEV of a singlet field $S$ via the
interaction $\lambda SH_d H_u$. For example, such a singlet coupling can be
enforced by a low energy $U(1)'$ gauge symmetry arising from a
high energy $E_6$ GUT group \cite{E6}. Within the class of $E_6$ models
there is a unique choice of Abelian gauge group, referred to as $U(1)_{N}$, which allows zero
charges for right-handed neutrinos. This choice of $U(1)_{N}$, which allows large
right-handed neutrino Majorana masses,
and hence a high scale see-saw mechanism,
defines the so-called Exceptional Supersymmetric
Standard Model (E$_6$SSM) \cite{King:2005jy,Athron:2009bs}.

In the E$_6$SSM, in order to cancel gauge anomalies involving $U(1)_{N}$,
the low energy (TeV scale) theory
must contain the matter content of three complete
27 representations of $E_6$ (minus the neutral
right-handed neutrinos which acquire intermediate scale masses).
It is clear that the E$_6$SSM predicts a rich spectrum of new states at the TeV scale
corresponding to the matter content of three 27 component families.
Since each 27 includes a pair of Higgs doublets plus a singlet,
the E$_6$SSM predicts in total three families of Higgs doublets and three families of Higgs singlets
\footnote{Each 27 component family also includes a pair of vector-like
charged $\pm 1/3$ coloured states $D,\bar{D}$ which are readily
produced at the LHC and provide a clear signature of the model.}.
The two Higgs doublets familiar from the
MSSM are denoted as $H_d$ and $H_u$,
while the two further replicas of these Higgs doublets predicted by the E$_6$SSM
are denoted as $H^d_1$, $H^u_1$ and $H^d_2$, $H^u_2$. Each
27 representation also contains a separate SM singlet, namely
the singlet $S$ whose VEV yields an effective $\mu$
term, plus two further copies of this singlet,
$S_1$ and $S_2$. In the E$_6$SSM the extra Higgs doublets,
$H^d_1$, $H^u_1$, $H^d_2$, $H^u_2$, and singlets, $S_1$,
$S_2$, are not supposed to develop VEVs and the scalar components
of these superfields are consequently called ``inert''.
From the perspective of dark matter, of particular interest are the fermionic partners of these
inert Higgs doublet and singlet
superfields, which we refer to as ``inert Higgsinos/singlinos''.
Such inert Higgsinos/singlinos will in
general mix with the other neutralinos and therefore change the
nature of lightest neutralino. If the LSP is the lightest
neutralino, identified as a WIMP CDM candidate, then the
calculation of the thermal relic density will necessarily be
affected by the presence of such inert Higgsinos/singlinos.

The purpose of this paper is to study neutralino dark matter in the presence of inert
Higgsinos/singlinos. As an example, we shall consider the
extended neutralino sector of the E$_6$SSM, which the includes three families
of Higgs doublet pairs, plus three singlets, plus a $Z'$, together
with their fermionic superpartners. The study here should be compared to that of
the USSM \cite{Kalinowski:2008iq} which, in addition to the
states of the MSSM, also includes a singlet, $S$, plus
plus a $Z'$, together with their fermionic superpartners, namely
the singlino $\tilde{S}$ and an extra gaugino $\tilde{B}'$.
In the USSM the neutralino LSP may have components
of the extra gaugino $\tilde{B}'$ and singlino $\tilde{S}$
in addition to the usual MSSM neutralino states, which can
have interesting consequences for the
calculation of the relic density $\Omega_\rr{CDM} h^2$.
In the present study we include all the above states of the USSM,
plus the extra inert Higgsino doublets predicted by the E$_6$SSM
but not included in the USSM, namely
$\tilde{H}^d_1$, $\tilde{H}^u_1$, $\tilde{H}^d_2$, $\tilde{H}^u_2$,
and the singlinos, $\tilde{S}_1$ and $\tilde{S}_2$, but we do not include the
corresponding inert scalars, which do not play a role in the heavy inert scalar limit.
We also do not include any of the exotic coloured states, $D$ and $\bar{D}$,
since in general we would not expect them to play a significant role in the
calculation of the dark matter relic abundance.

We shall study neutralino dark matter in the E$_6$SSM, as defined above,
both analytically and numerically, using {\tt MicrOMEGAs} \cite{o1}.
We find that results for the relic abundance
in the E$_6$SSM are radically different from those of both the MSSM
and the USSM. This is because the two families of inert doublet Higgsinos
and singlinos predicted by the E$_6$SSM provide an almost
decoupled neutralino sector with a naturally light LSP which can account for
the cold dark matter relic abundance independently of the
rest of the model. Typically the LSP will originate predominantly from
the neutralinos contained in the inert Higgsino/singlino families and
such an LSP will be able to account for the dark matter relic abundance
and satisfy current experimental data
\footnote{There is a lot of interest in the excess positron signal
that has been recently observed by the PAMELA, ATIC and Fermi collaboration (see e.g. \cite{pamela}). It has been speculated
that this could have been produced by annihilating dark matter in the galactic halo \cite{positronDM},
but it has also been suggested that the signal could be explained as coming from normal
astrophysical sources such as nearby pulsars \cite{pulsars}. In this paper we shall not try to interpret
these data as arising from neutralino dark matter, but instead we assume
some astrophysical explanation of the data.},
annihilating mainly through an s-channel $Z$-boson,
via its inert Higgsino doublet components which couple to the $Z$-boson.
This leads to a constraint that the LSP mass must exceed half the $Z$-boson
mass, to avoid the LEP constraints on the $Z$-boson width, which can be satisfied
providing that the ratio of the two usual
Higgs doublet VEVs ($\tan \beta$) is less than about 2.
Apart from the requirement $\tan \beta <2$,
the very stringent constraints on MSSM or USSM parameter space,
which come from requiring that the model explains the
relic density in terms of relic neutralinos, become completely relaxed,
since in the E$_6$SSM neutralino dark matter depends almost
exclusively on the parameters of the almost decoupled inert Higgsino sector.
We expect similar results to apply to any singlet-extended
SUSY model with an almost decoupled inert doublet Higgsino / Higgs singlino sector.

The remainder of the paper is organised as follows.
In Section~\ref{e6ssm} we briefly review the E$_6$SSM which provides
the motivation for including three families of
Higgs doublets and singlets.
In Section~\ref{inert} we discuss the inert Higgsino sector
of the E$_6$SSM and the effective model which we shall study, and highlight the most important
couplings for our analysis of the LSP dark matter relic density.
In Section~\ref{matrices} we display the complete neutralino and chargino mass
matrices of the considered model. In Section~\ref{analytical} we present some
analytical results which provide useful insight into the new inert
sector physics. These results are subsequently used to understand and interpret the
results of Section~\ref{results}, in which the results of a full numerical
dark matter relic density calculation using {\tt MicrOMEGAs} \cite{o1}
are presented. The paper is
concluded in Section~\ref{conclusion}.

\section{The E$_6$SSM}
\label{e6ssm}

One of the most important issues in models with additional Abelian
gauge symmetries is the cancellation of anomalies. In $E_6$
theories, if the surviving Abelian gauge group factor is a
subgroup of $E_6$ and the low energy spectrum constitutes complete
$27$ representations of $E_6$, then the anomalies are cancelled
automatically. In the E$_6$SSM the $27_i$ of $E_6$ containing the three quark and
lepton families decompose under the $SU(5)\times U(1)_{N}$
subgroup of $E_6$ as follows: \be 27_i\to
\ds\left(10,\,\ds{1}\right)_i+\left(5^{*},\,\ds{2}\right)_i
+\left(5^{*},\,-\ds{3}\right)_i +\ds\left(5,-\ds{2}\right)_i
+\left(1,\ds{5}\right)_i+\left(1,0\right)_i\,. \label{4} \ee
The first and second quantities in the brackets are the $SU(5)$
representation and extra $U(1)_{N}$ charge while $i$ is a family
index that runs from 1 to 3. From Eq.~(\ref{4}) we see that, in
order to cancel anomalies, the low energy (TeV scale) spectrum
must contain three extra copies of $5^*+5$ of $SU(5)$ in addition
to the three quark and lepton families in $5^*+10$. To be precise,
the ordinary SM families which contain the doublets of left-handed
quarks $Q_i$ and leptons $L_i$, right-handed up- and down-quarks
($u^c_i$ and $d^c_i$) as well as right-handed charged leptons, are
assigned to
$\left(10,\,\ds{1}\right)_i+\left(5^{*},\,\ds{2}\right)_i\,$.
Right-handed neutrinos $N^c_i$ should be associated with the last
term in Eq.~(\ref{4}), $\left(1,\,0\right)_i\,$. The next-to-last
term in Eq.~(\ref{4}), $\left(1,\,\ds{5}\right)_i\,$, represents
SM singlet fields $S_i$ which carry non-zero $U(1)_{N}$ charges
and therefore survive down to the EW scale. The three pairs of
$SU(2)$-doublets ($H^d_{i}$ and $H^u_{i}$) that are contained in
$\left(5^{*},\,-\ds{3}\right)_i$ and $\left(5,-\ds{2}\right)_i$
have the quantum numbers of Higgs doublets, and we shall identify
one of these pairs with the usual MSSM Higgs doublets, with the
other two pairs being inert Higgs doublets which do not get
VEVs. The other components of these $SU(5)$ multiplets form colour
triplets of exotic quarks, $D_i$ and $\overline{D}_i$, with electric
charges $-1/3$ and $+1/3$ respectively. The matter content and
correctly normalised Abelian charge assignment are in
Tab.~\ref{charges}.

\begin{table}[ht]
  \centering
  \begin{tabular}{|c||c|c|c|c|c|c|c|c|c|c|c|c|c|}
    \hline
 & $Q$ & $u^c$ & $d^c$ & $L$ & $e^c$ & $N^c$ & $S$ & $H_u$ & $H_d$ & $D$ &
 $\overline{D}$ & $H'$ & $\overline{H'}$ \\
 \hline
$\sqrt{\frac{5}{3}}Q^{Y}_i$
 & $\frac{1}{6}$ & $-\frac{2}{3}$ & $\frac{1}{3}$ & $-\frac{1}{2}$
& $1$ & $0$ & $0$ & $\frac{1}{2}$ & $-\frac{1}{2}$ & $-\frac{1}{3}$ &
 $\frac{1}{3}$ & $-\frac{1}{2}$ & $\frac{1}{2}$ \\
 \hline
$\sqrt{{40}}Q^{N}_i$
 & $1$ & $1$ & $2$ & $2$ & $1$ & $0$ & $5$ & $-2$ & $-3$ & $-2$ &
 $-3$ & $2$ & $-2$ \\
 \hline
  \end{tabular}
  \caption{\it\small The $U(1)_Y$ and $U(1)_{N}$ charges of matter fields in the
    E$_6$SSM, where $Q^{N}_i$ and $Q^{Y}_i$ are here defined with the correct
$E_6$ normalisation factor required for the RG analysis.}
  \label{charges}
\end{table}

We also require a further pair of superfields $H'$ and
$\overline{H'}$ with a mass term
$\mu' {H'}{\overline{H'}}$
from incomplete extra $27'$ and $\overline{27'}$
representations to survive to low energies to ensure gauge coupling
unification. Because $H'$ and $\overline{H'}$ originate from
$27'$ and $\overline{27'}$, these supermultiplets do not spoil anomaly
cancellation in the considered model.
Our analysis reveals that the unification of the gauge
couplings in the E$_6$SSM can be achieved for any phenomenologically
acceptable value of $\alpha_3(M_Z)$, consistent with the
measured low energy central value, unlike in the MSSM which requires
significantly higher values of $\alpha_3(M_Z)$, well above the central
measured value \cite{unif-e6ssm}
\footnote{The two superfields $H'$ and
$\overline{H'}$ may be removed from the spectrum,
thereby avoiding the $\mu'$ problem,
leading to unification at the string scale \cite{Howl:2008xz}.
However we shall not pursue this possibility in this paper.}.

Since right-handed neutrinos have zero charges
they can acquire very heavy Majorana masses. The heavy Majorana
right-handed neutrinos may decay into final states with lepton number
$L=\pm 1$, thereby creating a lepton asymmetry in the early Universe.
Because the Yukawa couplings of exotic particles are not constrained
by the neutrino oscillation data, substantial values of CP-violating lepton
asymmetries can be induced even for a
relatively small mass of the lightest right-handed neutrino ($M_1
\sim 10^6\,\mbox{GeV}$) so that successful thermal leptogenesis
may be achieved without encountering any gravitino problem \cite{King:2008qb}.

In $E_6$ models the renormalisable part of the superpotential arises
from the $27\times 27\times 27$ decomposition of the $E_6$ fundamental
representation. The most general renormalisable superpotential that
is allowed by the $E_6$ symmetry can be written in the following form:
\begin{eqnarray}
W_{E_6}&=&W_0+W_1+W_2\,, \label{cessm3}\\
W_0&=&\lambda_{ijk}S_i(H_{dj}H_{uk})+\kappa_{ijk}S_i(D_j\overline{D}_k)+h^N_{ijk}
N_i^c (H_{uj} L_k)+ h^U_{ijk} u^c_{i} (H_{uj}Q_k) \nonumber\\
&&+h^D_{ijk} d^c_i (H_{dj}Q_k) + h^E_{ijk} e^c_{i} (H_{dj} L_k)
\,,\label{W0} \\
W_1&=& g^Q_{ijk}D_{i} (Q_j Q_k)+g^{q}_{ijk}\overline{D}_i d^c_j u^c_k\,,\\
W_2&=& g^N_{ijk}N_i^c D_j d^c_k+g^E_{ijk} e^c_i D_j u^c_k+g^D_{ijk} (Q_i L_j) \overline{D}_k\,.
\end{eqnarray}

The superpotential of the E$_6$SSM clearly involves a lot of new Yukawa
couplings in comparison to the SM. In general these new interactions
violate baryon number conservation and induce non-diagonal flavour
transitions. To suppress baryon number violating and flavour changing
processes one can postulate a $Z^{H}_2$ symmetry under which all
superfields except one pair of $H^d_{i}$ and $H^u_{i}$ (say $H_d\equiv
H^d_{3}$ and $H_u\equiv H^u_{3}$) and one SM singlet field
($S\equiv S_3$) are odd. The $Z^{H}_2$ even Higgs doublets
then play the role of the conventional Higgs doublets which get
VEVs and are allowed to couple to the normal SM matter. Here we have
chosen the third generation to be even, so the inert superfields must
therefore belong to the first and second generations. The $Z^{H}_2$
symmetry then explains why the inert Higgs doublets and singlets do not get VEVs.

However, the $Z^{H}_2$ can only be approximate (otherwise the
exotics would not be able to decay).
To prevent rapid proton decay in the E$_6$SSM, a
generalised definition of $R$-parity should be used. We give two
examples of possible symmetries that can achieve this.
If $H^d_{i}$, $H^u_{i}$, $S_i$, $D_i$, $\overline{D}_i$
and the quark superfields ($Q_i$, $u^c_i$, $d^c_i$) are even under a
discrete $Z^L_2$ symmetry while the lepton superfields ($L_i$,
$e^c_i$, $N^c_i$) are odd (Model I)
then the allowed superpotential is invariant with respect to
a $U(1)_B$ global symmetry with the exotic $\overline{D}_i$ and
$D_i$ identified as diquark and anti-diquark, i.e.~$B_{D}=-2/3$ and
$B_{\overline{D}}=+2/3$. An alternative possibility is to assume that
the exotic quarks, $D_i$ and $\overline{D}_i$, as well as lepton
superfields, are all odd under $Z^B_2$ whereas the others remain
even. In this case (Model II) the $\overline{D}_i$ and
$D_i$ are leptoquarks \cite{King:2005jy}.

\section{The Inert Higgsino Couplings}
\label{inert}

The most important couplings in our analysis
are the trilinear couplings between
the three generations of up- and down-type Higgs doublets and Higgs SM
singlets contained in the superpotential of the E$_6$SSM in Eq.~(\ref{W0}),
\begin{equation}
\lambda_{ijk}S_iH_{dj}H_{uk} = \lambda_{ijk}(S_iH_{dj}^-H_{uk}^+ -
S_iH_{dj}^0H_{uk}^0). \label{lambda}
\end{equation}
The trilinear coupling tensor $\lambda_{ijk}$ in
Eq.~(\ref{lambda}) consists of 27 numbers, which play various roles.
The purely third family coupling
$\lambda_{333} \equiv \lambda$ is very important, because it is the
combination $\mu = \lambda s/\sqrt{2}$ that plays the role of
an effective $\mu$ term in this theory (where $s/\sqrt{2}$ is
the VEV of the third family singlet scalar $S_3 \equiv S$). Some
other neutralino mass terms, such as those involving $\tilde{S}$, are
also proportional to $\lambda$. The couplings of the inert (first
and second generation) Higgs doublet superfields to the third generation
Higgs singlet
superfield $\lambda_{3\alpha\beta} \equiv \lambda_{\alpha\beta}$ (where $\alpha,
\beta, \gamma$ index only the first and second generations)
directly contribute to neutralino and chargino mass terms for the
inert Higgsino doublets. $\lambda_{\alpha 3\beta} \equiv
f_{d\alpha\beta}$ and $\lambda_{\alpha\beta 3} \equiv f_{u\alpha\beta}$
directly contribute to neutralino mass terms involving an inert doublet Higgsino
and singlino.

The 13 Higgs trilinear couplings mentioned thus far are the only
couplings that obey the proposed $Z_2^H$ symmetry. This
symmetry (under which all superfields other than the third
generation $H_d$, $H_u$ and $S$ are odd) is proposed in order to
prevent flavour changing neutral currents in the SM matter sector
by eliminating non-diagonal flavour transitions. There is,
however, no specific reason to suspect that it is respected by the
$\lambda_{ijk}$ couplings or by superpotential couplings involving
the exotic quarks. Indeed, if $Z_2^H$ is respected by the
latter then the lightest exotic quark state(s) would be stable.
This would presumably lead to a relic density of heavy exotic
quark states inconsistent with observation. If $\lambda_{ijk}$
obeyed $Z_2^H$ exactly then, as we will see below, the
neutralino mass matrix (and also the chargino mass matrix) would be
decoupled into two independent systems and the lightest from each sector
would be stable. We
shall refer to the $Z_2^H$ breaking couplings involving two
third generation superfields as $\lambda_{3\alpha 3} \equiv
x_{d\alpha}$, $\lambda_{33\alpha} \equiv x_{u\alpha}$ and
$\lambda_{\alpha 33} \equiv z_\alpha$. The notation for the
$\lambda_{ijk}$ couplings used in this paper are compiled in Tab.~\ref{couplings}.

\begin{table}[ht]
  \centering
\begin{tabular}{|c||c|c|c|c|c|c|c|}
\hline
$\lambda_{ijk}$ & $\lambda$ & $\lambda_{\alpha\beta}$ & $f_{d\alpha\beta}$ & $f_{u\alpha\beta}$ & $x_{d\alpha}$ & $x_{u\alpha}$ & $z_\alpha$ \\
\hline
$ijk$ & $333$ & $3\alpha\beta$ & $\alpha 3\beta$ & $\alpha\beta 3$ & $33\alpha$ & $3\alpha 3$ & $\alpha 33$ \\
\hline
  \end{tabular}
  \caption{\it\small The notation for the
$\lambda_{ijk}$ couplings.}
  \label{couplings}
\end{table}

The remaining 8 $Z_2^H$ breaking couplings
$\lambda_{\alpha\beta\gamma}$ are of less importance. As long as
only the third generation Higgs doublets and singlet acquire VEVs
then these couplings do not appear in the neutralino or chargino
mass matrices. Additionally, they only appear in Feynman
rules that involve the inert Higgs scalars and we assume
that these are given soft SUSY
breaking masses that are heavy enough such that these particles do
not contribute to any processes relevant for the current study.

As a final note, one could perhaps argue that these couplings should
be arranged to help ensure that only the third generation
singlet scalar radiatively acquires a VEV. However, as the
contributions to the running of the singlet scalar square masses
could be coming mostly from the heavy exotic quarks, there is
little reason to impose any constraints from such considerations
on the $\lambda_{ijk}$ couplings.

\section{The Neutralino and Chargino Mass Matrices}
\label{matrices}

In the MSSM there are four neutralino interaction states, the
neutral wino, the bino and the two Higgsinos. In the USSM, two
extra states are added, the singlino and the bino$'$. In
the conventional USSM basis
\begin{equation}
\tilde{\chi}_\rr{int}^0 = (\begin{array}{cccc|cc} \tilde{B} & \tilde{W}^3 & \tilde{H}_d^0 &
\tilde{H}_u^0 & \tilde{S} & \tilde{B}^\prime \end{array})^\rr{T}
\end{equation}
and neglecting bino-bino$'$ mixing (as justified in Ref.~\cite{Kalinowski:2008iq})
the USSM neutralino mass matrix is then
\begin{equation}\small
M^n_\rr{USSM} = \left( \begin{array}{cccc|cc} M_1 & 0 & -m_Zs_Wc_\beta & m_Zs_Ws_\beta & 0 & 0\\
0 & M_2 & m_Zc_Wc_\beta & -m_Zc_Ws_\beta & 0 & 0\\
-m_Zs_Wc_\beta & m_Zc_Wc_\beta & 0 & -\mu & -\mu_ss_\beta & g_1^\prime vc_\beta Q_d^N \\
m_Zs_Ws_\beta & -m_Zc_Ws_\beta & -\mu & 0 & -\mu_sc_\beta &
g_1^\prime vs_\beta Q_u^N \\ \hline
0 & 0 & -\mu_ss_\beta & -\mu_sc_\beta & 0 & g_1^\prime sQ_s^N \\
0 & 0 & g_1^\prime vc_\beta Q_d^N & g_1^\prime vs_\beta
Q_u^N & g_1^\prime sQ_s^N & M_1^\prime \end{array}
\right),
\label{USSM}
\end{equation}
where $M_1$, $M_2$ and $M_1^\prime$ are the soft gaugino masses,
$\mu_s = \lambda v/\sqrt{2}$, $\<H_d\> = v \cos \beta /\sqrt{2}$ and
$\<H_u\> = v \sin \beta/\sqrt{2}$. In the E$_6$SSM this is
extended. We take the full basis of neutralino interaction states to be
\begin{equation}
\tilde{\chi}_\rr{int}^0 = (\begin{array}{cccc|cc|ccc|ccc}
\tilde{B} & \tilde{W}^3 & \tilde{H}_d^0 & \tilde{H}_u^0 &
\tilde{S} & \tilde{B}^\prime & \tilde{H}_{d2}^0 & \tilde{H}_{u2}^0
& \tilde{S}_2 & \tilde{H}_{d1}^0 & \tilde{H}_{u1}^0 & \tilde{S}_1
\end{array})^\rr{T}.
\end{equation}
The first four states are the MSSM interaction states, the $\tilde{S}$
and $\tilde{B}^\prime$ are the extra states added in the USSM and
the final six states are the extra inert doublet Higgsinos and Higgs singlinos
that come with the full E$_6$SSM model.
Under the assumption that only the third generation Higgs doublets
and singlet acquire VEVs the full Majorana mass matrix is then
\begin{equation}
M^n_{\rr{E}_6\rr{SSM}} = \left( \begin{array}{ccc} M^n_\rr{USSM} & B_2 & B_1\\
B_2^\rr{T} & A_{22} & A_{21}\\
B_1^\rr{T} & A_{21}^\rr{T} & A_{11}\end{array} \right),\label{nmm}
\end{equation}
where the sub-matrices involving the inert interaction states are given by
\begin{equation}
A_{\alpha\beta} = -\frac{1}{\sqrt{2}} \left( \begin{array}{ccc}
0 & \lambda_{\alpha\beta} s & f_{u\beta\alpha}v\sin \beta\\
\lambda_{\beta\alpha} s & 0 & f_{d\beta\alpha}v\cos \beta\\
f_{u\alpha\beta}v\sin \beta & f_{d\alpha\beta}v\cos \beta & 0 \end{array}
\right),\label{AA}
\end{equation}
and the $Z_2^H$ breaking sub-matrices by
\begin{equation}
B_\alpha = -\frac{1}{\sqrt{2}}
\left( \begin{array}{ccc} 0 & 0 & 0 \\ 0 & 0 & 0 \\
0 & x_{d\alpha}s & z_\alpha v\sin \beta \\
x_{u\alpha}s & 0 & z_\alpha v\cos \beta \\
x_{u\alpha}v\sin \beta &
x_{d\alpha}v\cos \beta & 0 \\ 0 & 0 & 0
\end{array} \right).
\end{equation}

Similarly we take our basis of chargino interaction states to be
\begin{equation}
\tilde{\chi}_\rr{int}^{\pm} = \left( \begin{array}{c}
\tilde{\chi}_\rr{int}^+ \\ \tilde{\chi}_\rr{int}^- \end{array}
\right ), \nonumber\
\end{equation}
where
\begin{equation}
\begin{array}{ccc}
\tilde{\chi}_{\rr{int}}^+ = \left( \begin{array}{c} \tilde{W^+} \\
\tilde{H}_u^+ \\ \tilde{H}_{u2}^+ \\ \tilde{H}_{u1}^+ \end{array}
\right) & \rr{and} & \tilde{\chi}_{\rr{int}}^- = \left(
\begin{array}{c} \tilde{W^-} \\ \tilde{H}_d^- \\ \tilde{H}_{d2}^-
\\ \tilde{H}_{d1}^- \end{array} \right).
\end{array}
\end{equation}
The corresponding mass matrix is then
\begin{equation}
M^c_{\rr{E}_6\rr{SSM}} = \left( \begin{array}{cc} & C^\rr{T} \\ C
\end{array} \right), \nonumber
\end{equation}
where
\begin{equation}
C = \left( \begin{array}{cccc} M_2 & \sqrt{2}m_W\sin \beta & 0 & 0\\
\sqrt{2}m_W\cos \beta & \mu & \frac{1}{\sqrt{2}}x_{d2}s &
\frac{1}{\sqrt{2}}x_{d1}s \\
0 & \frac{1}{\sqrt{2}}x_{u2}s & \frac{1}{\sqrt{2}}\lambda_{22}s & \frac{1}{\sqrt{2}}\lambda_{21}s \\
0 & \frac{1}{\sqrt{2}}x_{u1}s & \frac{1}{\sqrt{2}}\lambda_{12}s &
\frac{1}{\sqrt{2}}\lambda_{11}s \end{array} \right).
\label{cmm}
\end{equation}

It is clear that a generic feature of the E$_6$SSM is that the LSP is
usually (naturally) composed mainly of inert singlino and ends
up being typically very light. One can see this by inspecting the
new sector blocks of the extended neutralino mass matrix in Eq.~(\ref{nmm}),
such as $A_{11}$, and assuming a hierarchy of the form
$\lambda_{\alpha\beta} s \gg f_{(u,d)\alpha\beta}v$.
This is a natural assumption since we
already require that $s \gg v$ in order to satisfy the current experimental limit
on the $Z'$ mass of around 1 TeV \cite{Aaltonen:2008ah}, as discussed in Ref.~\cite{Athron:2009bs}.

For both the neutralinos and the charginos we see that if the
$Z_2^H$ breaking couplings are exactly zero then the new part
of the E$_6$SSM mass matrix becomes decoupled from the USSM mass
matrix. However, although approximate decoupling is expected, exact decoupling is not,
and will therefore not be considered.

\section{Analytical Discussion}
\label{analytical}

According to standard cosmology,
at some time in the past, before Big Bang Nucleosynthesis (BBN), the LSP
would have decoupled from equilibrium with other
species still in equilibrium with the photon. This decoupling from
chemical equilibrium would have happened roughly when the particle's inelastic
interaction rate (maintaining chemical equilibrium) became less
than the expansion rate of the universe $H = \dot{a}/a$, where $a$
is the scale factor of the universe. When
such a chemical ``freeze-out'' occurs the number density of the
frozen out species (the LSP here) typically remains much larger
than it would have been if the species had remained in chemical
equilibrium as the universe cooled. From this point onwards it is
approximately just the number density at freeze-out
that determines the relic density of the stable particle today.
Generally the larger a stable relic's annihilation and co-annihilation
cross-sections would have been before freeze-out, the lower its relic density
in the universe would be today \cite{d1}.

In order for such a relic particle to be ``cold'' (as in ``cold dark
matter'') the freeze-out temperature must be much less than the mass
of the particle, such that the particle was non-relativistic at
freeze-out. The measured value used for the total present day
cold dark matter relic density is $\Omega_\rr{CDM}h^2 = 0.1099 \pm
0.0062$ \cite{p1}. If a theory predicts a greater relic density
of dark matter than this then it is ruled out, assuming standard
pre-BBN cosmology. A theory that predicts less dark matter cannot in the
same way be ruled out, but if the theory is supposed to be the low
energy effective theory of the complete theory that describes the
universe then it should
account for all of the observed dark matter. The LSP relic density
calculation has already been widely studied in the MSSM \cite{all}
and especially in the constrained MSSM \cite{Kane:1993td}.

It will be useful to get some analytical understanding of the calculation of the
relic abundance coming from the new
neutralino/chargino physics of the E$_6$SSM before looking at
the results of the full numerical simulation. To this end,
in this section, we consider just
one inert Higgs family consisting of two inert Higgs doublets and one inert Higgs singlet,
which we shall label as the first generation. We shall assume that
the $Z_2^H$ breaking couplings of the first (inert) Higgs generation to the
third (conventional) Higgs generation are large enough to allow the heavier
states of the USSM to decay into the LSP, formed mostly from inert states,
but also small enough such that we can consider the
inert Higgsinos to be approximately decoupled from the rest of the
neutralino mass matrix for the purposes of obtaining an
analytical estimate of the mass eigenstates.
This amounts to considering the single block $A_{11}$
of the extended neutralino mass matrix in Eq.~(\ref{nmm})
and ignoring the rest.
We emphasise that this is for the purposes of the simple analytical estimates
in this section only and that in the next section we shall perform a full numerical
analysis without any approximation.

\subsection{Inert Neutralino Masses and Mixing for One Family}

Within the first generation we use the basis
\begin{equation}
\tilde{\chi}_{\rr{int}}^0 = (\begin{array}{ccc} \tilde{H}_{d1}^0 &
\tilde{H}_{u1}^0 & \tilde{S}_1 \end{array})^\rr{T}
\end{equation}
and the neutralino mass matrix is then, from Eq.~(\ref{AA}),
\begin{equation}
A_{11} \equiv A = -\frac{1}{\sqrt{2}} \left( \begin{array}{ccc} 0 &
\lambda^\prime s & f_uv\sin \beta \\ \lambda^\prime s & 0 &
f_dv\cos \beta \\ f_uv\sin \beta & f_dv\cos \beta & 0 \end{array}\right),
\end{equation}
where $\lambda^\prime = \lambda_{11} \equiv \lambda_{311}$, $f_d =
f_{d11} \equiv \lambda_{131}$ and $f_u = f_{u11} \equiv \lambda_{113}$. As
discussed earlier, it is natural to assume that $\lambda^\prime s
\gg fv$ and this will lead to a light, mostly
first-generation-singlino lightest neutralino.

Finding the mass eigenvalues of the matrix $A$ involves solving a
reduced cubic equation. Doing an expansion in $fv/\lambda^\prime s$
the three neutralino masses from the first generation are
\begin{eqnarray}
m_1 &=& \frac{1}{\sqrt{2}} \frac{f_df_u}{\lambda^\prime} \frac{v^2}{s} \sin(2\beta)
+ \cdots,\label{sin2beta} \label{m1} \\
m_2 &=& \frac{\lambda^\prime s}{\sqrt{2}} - \frac{m_1}{2} + \cdots, \\
m_3 &=& -\frac{\lambda^\prime s}{\sqrt{2}} - \frac{m_1}{2} + \cdots.
\end{eqnarray}
The lightest state is mostly singlino (as we will confirm below)
and the two heavier states are nearly mass degenerate, split by
the LSP mass. At $\beta = 0$ or $\pi/2$ the lightest neutralino
becomes massless. This is when only one of the third generation
conventional Higgs doublets has a VEV. The LSP, even if very
weakly interacting, must be heavier than a few MeV so that it
would not contribute to the expansion rate prior to
nucleosynthesis, changing nuclear abundances \cite{King:2005jy}.

We shall define the neutralino mixing matrix $N$ by
\begin{equation}
N_i^a M^{ab} N_j^b = m_i \delta_{ij} \mbox{ no sum on } i.
\end{equation}
The lightest state is then made up of the following superposition
of interaction states:
\begin{equation}
\tilde{\chi}_1^0 = N_1^1 \tilde{H}_{d1}^0 + N_1^2 \tilde{H}_{u1}^0
+ N_1^3 \tilde{S}_1.
\end{equation}
Again expanding in $fv/\lambda^\prime s$ we have
\begin{equation}
N_1 = \left( \begin{array}{c} -\frac{f_dv}{\lambda^\prime s} \cos \beta + \cdots \\ \\ -\frac{f_uv}{\lambda^\prime s} \sin \beta + \cdots \\ \\
1 - \frac{1}{2} \left(\frac{v}{\lambda^\prime s}\right)^2
\left[f_d^2\cos^2(\beta) + f_u^2\sin^2(\beta)\right] + \cdots\label{N1}
\end{array} \right),
\end{equation}
confirming that the LSP is mostly singlino in this limit. The
other eigenvectors, which determine the composition of neutralinos 2 and 3, are
\begin{equation}
N_i = \sqrt{\frac{1}{a_i^2 + b_i^2 + \cdots}} \left(
\begin{array}{c} a_i \\ b_i \\ 1 \end{array} \right),
\end{equation}
where
\begin{eqnarray}
-b_2 = a_2 &=& \frac{\lambda^\prime s}{v} \left[f_d\cos \beta - f_u\sin \beta\right]^{-1} + \cdots, \\ \nonumber \\
b_3 = a_3 &=& \frac{\lambda^\prime s}{v} \left[f_d\cos \beta + f_u\sin \beta\right]^{-1} + \cdots.
\end{eqnarray}
Note that $a,b \gg 1$ and that $a_2$ and $b_2$ flip sign at
$f_d\cos \beta = f_u\sin \beta$ whereas $a_3$ and $b_3$ are
always positive. Very approximately these eigenvectors are then
\begin{eqnarray}
N_2 &=& \frac{1}{\sqrt{2}} \left( \begin{array}{c} -1 \\ 1 \\ 0 \end{array} \right) \rr{sign}(f_us_\beta - f_dc_\beta), \\ \nonumber \\
N_3 &=& \frac{1}{\sqrt{2}} \left( \begin{array}{c} 1 \\ 1 \\ 0
\end{array} \right).
\end{eqnarray}

Under the assumptions of this section the lightest chargino is
simply the first generation charged Higgsino with a mass $m_c =
\lambda^\prime s/\sqrt{2}$.

\subsection{Annihilation Channels}

From Eq.~(\ref{m1}) it is seen that the LSP mass $m_1$ is proportional to $v^2/s$
and so is naturally small since $v\ll s$. To understand this, recall that
$Z$-$Z^\prime$ mixing leads to two mass eigenstates, $Z_2 \sim Z'$ and $Z_1 \sim Z$,
and limits on $Z$-$Z^\prime$ mixing and on the $Z_2$ mass place lower limits on $s$,
with $v\ll s$ being always satisfied. For example, when $s=3000$ GeV
the $Z_2$ mass is about 1100 GeV and $v^2/s \approx 20$ GeV. The LSP mass further decreases
as $s$ becomes larger in the considered limit. In practice, it is quite difficult to arrange
the LSP mass to exceed about 100 GeV.


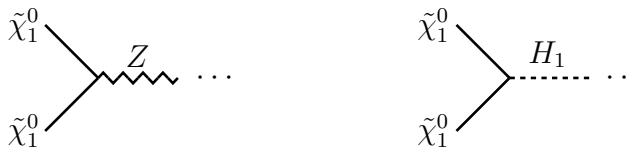
\begin{figure}[h]
\begin{center}
\begin{picture}(235,60)(-15,-30)
\SetWidth{1}
\Text(3,20)[r]{$\tilde{\chi}_1^0$}
\Text(3,-20)[r]{$\tilde{\chi}_1^0$}
\Line(5,20)(25,0)
\Line(5,-20)(25,0)
\ZigZag(25,0)(55,0){2}{4}
\Text(40,4)[b]{$Z$}
\Text(62,0)[l]{$\cdots$}
\Text(158,20)[r]{$\tilde{\chi}_1^0$}
\Text(158,-20)[r]{$\tilde{\chi}_1^0$}
\Line(160,20)(180,0)
\Line(160,-20)(180,0)
\DashLine(180,0)(210,0){2}
\Text(195,4)[b]{$H_1$}
\Text(217,0)[l]{$\cdots$}
\end{picture}
\end{center}
\caption{\it\small s-channel LSP annihilation diagrams.}\label{sChan}
\end{figure}

In view of the above discussion the LSP is expected to be relatively light,
and so we begin by looking at s-channel annihilation,
which can result in lighter mass final states. The most important
diagrams are shown in Fig.~\ref{sChan} and it will turn out that the
most important of these annihilations have a $Z$-boson in the s-channel
(or strictly speaking $Z_1 \sim Z$).
The $\tilde{\chi}^0_1\tilde{\chi}^0_1Z$ gauge coupling in this diagram
is suppressed by a factor of
\begin{equation}
\frac{1}{2} \left(\frac{v}{\lambda^\prime s}\right)^2 \left[f_u^2\sin^2(\beta) -
f_d^2\cos^2(\beta)\right] + \cdots \nonumber
\end{equation}
under the assumptions of this section, since the LSP only couples
through its small Higgsino components. This coupling vanishes
completely at $f_d\cos \beta = f_u\sin \beta$, which is when the
LSP contains a completely symmetric combination of
$\tilde{H}_{d1}^0$ and $\tilde{H}_{u1}^0$. In the MSSM a Higgsino-like
LSP is typically such a symmetric combination of up- and
down-type Higgsino and therefore does not couple very strongly to
the $Z$-boson. In this model, however, the LSP is
unlikely to have very similar admixtures of $\tilde{H}_{d1}^0$ and
$\tilde{H}_{u1}^0$.

Full gauge coupling strength s-channel $Z$-boson
annihilations tend to leave a relic density that is too low to
account to for the observed dark matter,
but in this model the coupling of the
mostly singlino LSP to the $Z$-boson is typically suppressed,
as it only couples through its doublet Higgsino admixture, leading to
an increased relic density if this is the dominant annihilation
channel. As $\lambda' s$ decreases,
the proportion of the LSP that is made up of inert doublet Higgsino,
rather than inert singlino, increases. This can be seen in Eq.~(\ref{N1}).
This then increases the strength of the
overall $\tilde{\chi}^0_1\tilde{\chi}^0_1Z$ coupling. The inclusive cross-section for
s-channel annihilation through a $Z$-boson is therefore highly dependent
on $\lambda' s$, which affects both the coupling and the LSP mass $m_1$.
The effect of independently increasing the coupling is always to increase the cross-section,
but the effect of independently increasing the LSP mass can be to either increase or decrease the
cross-section, depending on which side of the $Z$-boson resonance it is on.
In the considered limit both the mass and coupling are proportional to $1/(\lambda^\prime s)^2$,
and the annihilation cross-section is given by,
\begin{equation}
\sigma(\tilde{\chi}_1^0\tilde{\chi}_1^0 \rightarrow Z*
\rightarrow \mbox{ anything}) \propto
\left(\frac{1}{\lambda^\prime s}\right)^4
\left(\frac{1}{m_Z^2-(2m_1)^2}\right)^2
\left(f_u^2s_\beta^2-f_d^2c_\beta^2\right)^2.
\end{equation}
The s-channel annihilation through the
lightest Higgs boson will also become important if the LSPs are on resonance
in the relic density calculation.

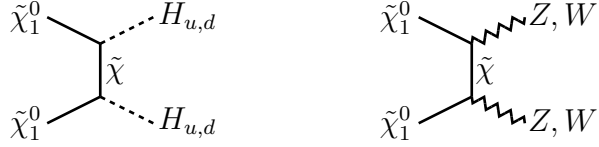
\begin{figure}
\begin{center}
\begin{picture}(210,60)(-10,-30)
\SetWidth{1}
\Text(8,20)[r]{$\tilde{\chi}_1^0$}
\Text(8,-20)[r]{$\tilde{\chi}_1^0$}
\Line(10,20)(30,10)
\Line(10,-20)(30,-10)
\Line(30,10)(30,-10)
\DashLine(30,10)(50,20){2}
\DashLine(30,-10)(50,-20){2}
\Text(32,0)[l]{$\tilde{\chi}$}
\Text(52,20)[l]{$H_{u,d}$}
\Text(52,-20)[l]{$H_{u,d}$}
\Text(148,20)[r]{$\tilde{\chi}_1^0$}
\Text(148,-20)[r]{$\tilde{\chi}_1^0$}
\Line(150,20)(170,10)
\Line(150,-20)(170,-10)
\Line(170,10)(170,-10)
\ZigZag(170,10)(190,20){2}{4}
\ZigZag(170,-10)(190,-20){-2}{4}
\Text(172,0)[l]{$\tilde{\chi}$}
\Text(192,20)[l]{$Z,W$}
\Text(192,-20)[l]{$Z,W$}
\end{picture}
\end{center}
\caption{\it\small t-channel LSP annihilation diagrams.}\label{tChan}
\end{figure}

The most important of the potential t-channel processes are shown in Fig.~\ref{tChan}.
In practice these channels will not play a significant role compared to the
s-channel annihilations considered previously, but we discuss them for completeness.
The t-channel particle for these
processes is one of the neutralinos or the chargino of the first
generation (for producing neutral Higgs scalars / $Z$-bosons or
charged Higgs scalars / $W$-bosons respectively). In the first
diagram, t-channel annihilation to conventional third generation
Higgs scalars, the couplings are just $f$ couplings of the first
generation and appropriate mixing matrix elements.  With the
chargino or with neutralino 2 or 3 in the t-channel the diagram is
approximately inert singlinos annihilating with an inert doublet
Higgsino in the t-channel and the couplings are approximately just
$f_d$ and $f_u$ for producing $H_d$ and $H_u$ interaction states respectively.
The LSP mass is smaller than the other masses by a
factor of order $v^2/s^2$. With another LSP in the t-channel the
first diagram therefore receives an enhancement of order $s^2/v^2$
for the t-channel propagator at low momentum, but has a
suppression of order $v^2/s^2$ in the couplings due to the LSP
only containing doublet type first generation Higgsinos with
amplitudes of order $v/s$.

The second diagram in Fig.~\ref{tChan} represents annihilation to massive gauge bosons. To
very good approximation these bosons only couple to weak isospin
doublets and not to SM singlets (since $Z$-$Z'$ mixing must be small).
These diagrams are therefore
suppressed by order $v^2/s^2$ in the couplings even with a
chargino or with neutralino 2 or 3 in the t-channel. On top
of this suppression these diagrams also receive an additional
suppression of order $v^2/s^2$ in the couplings, but an enhancement
of order $s^2/v^2$ in the propagator when the t-channel contains
the LSP. Although this second type of diagram is suppressed
relative to the first (assuming $v^2/s^2 \ll f$) it has a greater
chance of being kinematically allowed. As previously stated,
inert scalar Higgs-bosons are assumed heavy and annihilation to
these particles is not considered.

\section{Numerical Analysis}
\label{results}

We now turn to the full model, in which the LSP is determined from the neutralino mass matrix in
Eq.~(\ref{nmm}) where there are two copies of the family considered in
the previous section as well as 6 unknown mixing parameters between the
two families. In general, after rotation to the mass
eigenstates, we expect that two states are much lighter than the rest,
both inert-singlino-like in the $\lambda^\prime s \gg fv$
limit \footnote{An exception to this is in the large $M'_1$ limit
in which the LSP could originate from the lower block of the USSM neutralino mass
matrix in Eq.~(\ref{USSM}) due to a mini see-saw mechanism as discussed in Ref.~\cite{Kalinowski:2008iq}.}.

In this section we use numerical methods to
predict the relic density. We first diagonalize
the neutralino, chargino and Higgs scalar mass
matrices numerically. The 1-loop USSM Higgs mass corrections from top and
stop loops are implemented from Ref.~\cite{c1}. Corrections from
exotic quark and squark loops are not included in our analysis,
as these have been
shown to be small \cite{King:2005jy}, and CP violation is not considered.
Having done this {\tt MicrOMEGAs 2.2} \cite{o1} is then used to numerically compute the
present day relic density, including the relevant (co-)annihilation channel
cross-sections and the LSP freeze-out temperature. {\tt MicrOMEGAs}
achieves this by calculating all of the relevant tree-level
Feynman diagrams using {\tt CalcHEP}. The {\tt CalcHEP} model files for the
considered model are generated using
{\tt LanHEP} \cite{o2}. The {\tt MicrOMEGAs} relic density calculation
assumes standard cosmology in which the LSP was in equilibrium
with the photon at some time in the past.

\subsection{The Parameter Space of the Model}

Motivated by the running of the gauge couplings from the GUT scale,
we assume that the GUT normalised couplings of the two $U(1)$
gauge groups, $U(1)_Y$ and $U(1)_N$, are equal and that the mixing
between the two groups is negligible, giving $g_1^\prime \approx
0.46$. The free parameters are then
the trilinear Higgs couplings $\lambda_{ijk}$, the singlet VEV
$s$, $\tan \beta$, the soft $\lambda_{333}$ coupling $A_\lambda$
and the soft gaugino masses. It will turn out that the soft
gaugino masses usually have little effect on the dark matter
physics. One can see this by observing the neutralino mass
matrix, Eq.~(\ref{nmm}), where
the USSM terms coming from the soft gaugino masses do not directly
mix with terms from the new E$_6$SSM inert sector. The scalar
Higgs doublet and singlet soft SUSY breaking masses
are determined from the scalar potential
minimalisation conditions given
$s$, $v$, $\tan \beta$ and $A_\lambda$.
The regular squark and slepton sectors
as well as the potential issue of mixing between the two $U(1)$
gauge groups are the same as in the USSM \cite{Kalinowski:2008iq}.

In the following analysis we shall choose $s=3000$~GeV and
$\mu=400$~GeV which gives $\lambda=2\sqrt{2}\,/15 \approx 0.19$
and makes the $Z_2$ (i.e.~$Z'$) mass about 1100~GeV. Although much of the physics
is highly dependent on $s$, this specific choice of $s$ does not
limit the generality of the results obtained. This is explained below.
We also choose $M_1 = M_1^\prime = M_2/2 = 250$ GeV. These relations between the
soft gaugino masses are motivated by their running from high
scale, but the value is not. In this analysis the squarks and sleptons will not
play a significant role in the calculation of dark matter relic abundance
since the LSP will always be much lighter.
We choose equal soft sfermion masses to be $M_S = 800$~GeV and the
stop mixing parameter, $X_t = A_t - \mu\cot(\beta)$, to be $X_t =
\sqrt{6}\,M_S$ as in Ref.~\cite{King:2005jy}. This results in a lightest
CP-even Higgs mass in excess of 114~GeV for all parameter
space considered below. The soft $\lambda$ coupling $A_\lambda$
is set by choosing the pseudo-scalar Higgs mass $m_A$. We choose
$m_A = 500$~GeV.

We initially assume the $Z_2^H$ breaking $\lambda_{ijk}$
couplings to be small (0.01) for the following analysis. The main
properties of the physics can then be seen by varying three
parameters $\lambda^\prime=\lambda_{22}=\lambda_{11}$,
$f=f_{d22}=f_{u22}=f_{d11}=f_{u11}$ and $\tan \beta$. The first
and second generation mixing couplings are set to
$\lambda_{21,12}=\epsilon\lambda^\prime$ and
$f_{(d,u)(21,12)}=\epsilon f$. Assuming this parameter choice the
sub-matrices of the neutralino mass matrix Eq.~(\ref{AA}) become
\begin{eqnarray}
A_{22} &=& A_{11} = -\frac{1}{\sqrt{2}} \left( \begin{array}{ccc}
0 & \lambda^\prime s & fv\sin \beta\\
\lambda^\prime s & 0 & fv\cos \beta\\
fv\sin \beta & fv\cos \beta & 0 \end{array} \right)\,. \label{param1} \\
A_{21} &=& \epsilon A_{22}. \label{param2}
\end{eqnarray}
This simple parametrisation is sufficient for
illustrating the generic properties of the physics. Deviations
from this parametrisation are discussed afterwards.

Note that the analytical results of the previous section provide an essential context
in which
to understand the numerical results of this section.
According to the above parametrization, the two generations
are approximately degenerate and the mixing terms are not too large. In this
case the LSP and the second lightest neutralino will each contain approximately
equal contributions from each generation.

Finally, it is worth remarking that, assuming the above parametrisation, the effect
on the neutralino and chargino inert sectors of changing $s$ is
simply equivalent to that of changing $\lambda^\prime$
(although the  $Z'$ mass will depend on $s$). This means that the
following results are also applicable for other experimentally
consistent values of $s$, scaled by $\lambda^\prime$.

\subsection{Neutralino and Chargino Spectra}

\begin{figure}[t]
\begin{center}
\includegraphics[width=150mm]{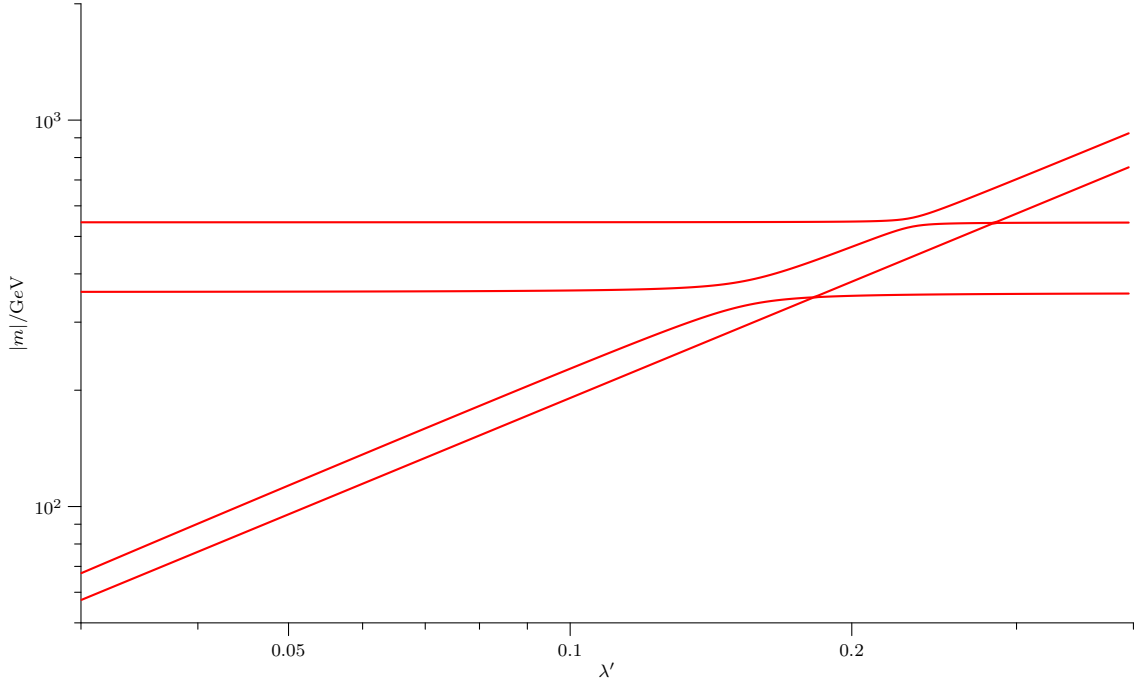}
\end{center}
\caption{\it\small Chargino masses (magnitude only) against
$\lambda^\prime$ with $f=1$, $\epsilon = 0.1$, $\tan \beta =
1.5$, $s=3000$~GeV and $Z_2^H$ breaking $\lambda_{ijk}$ couplings set to
0.01.}\label{specC}
\end{figure}

\begin{figure}[t]
\begin{center}
\includegraphics[width=150mm]{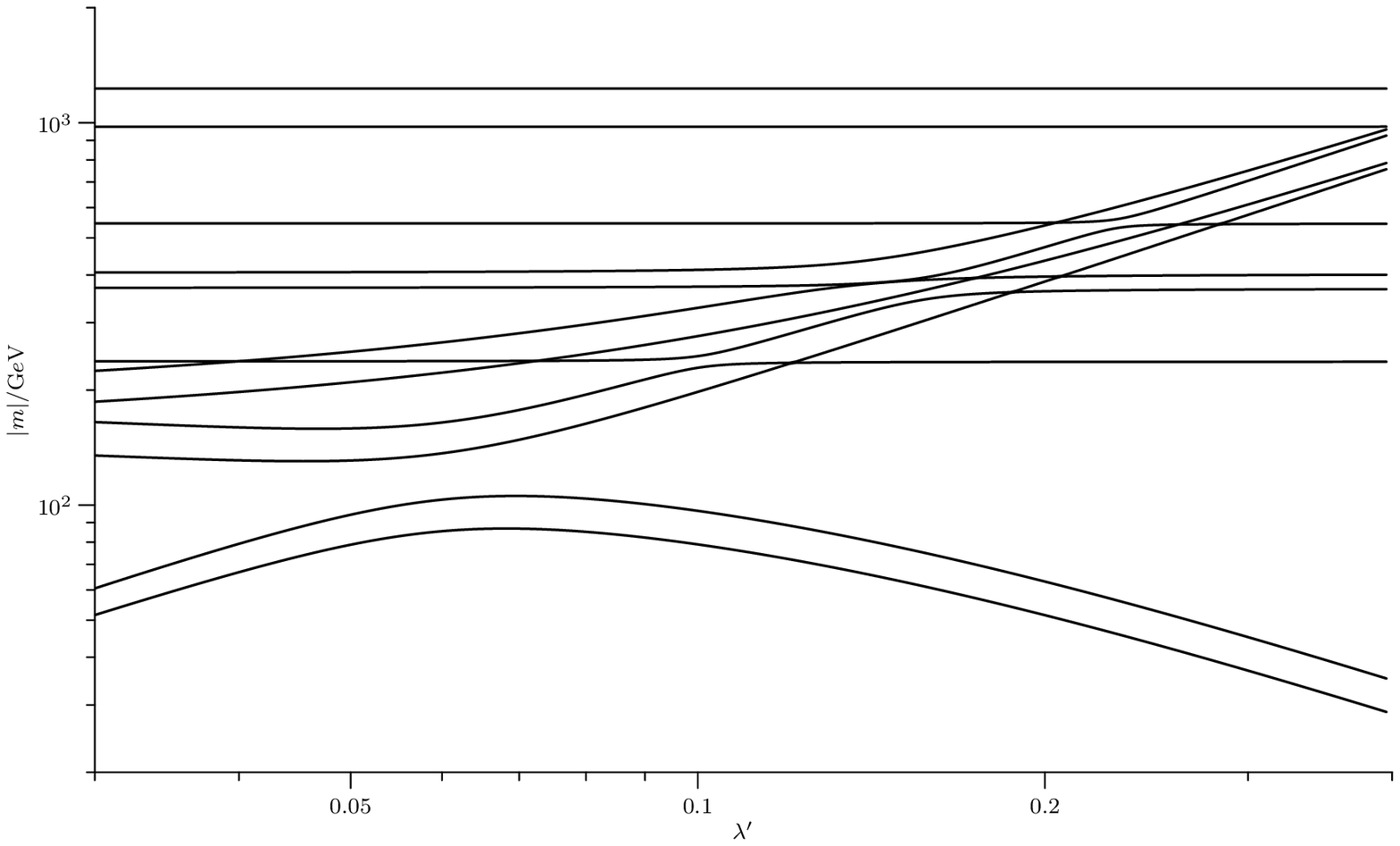}\end{center}
\caption{\it\small Neutralino masses (magnitude only) against $\lambda^\prime$
with $f=1$, $\epsilon = 0.1$, $\tan \beta = 1.5$, $s=3000$~GeV and
$Z_2^H$ breaking $\lambda_{ijk}$ couplings set to
0.01.}\label{specN}
\end{figure}

\begin{figure}[t]
\begin{center}
\includegraphics[width=150mm]{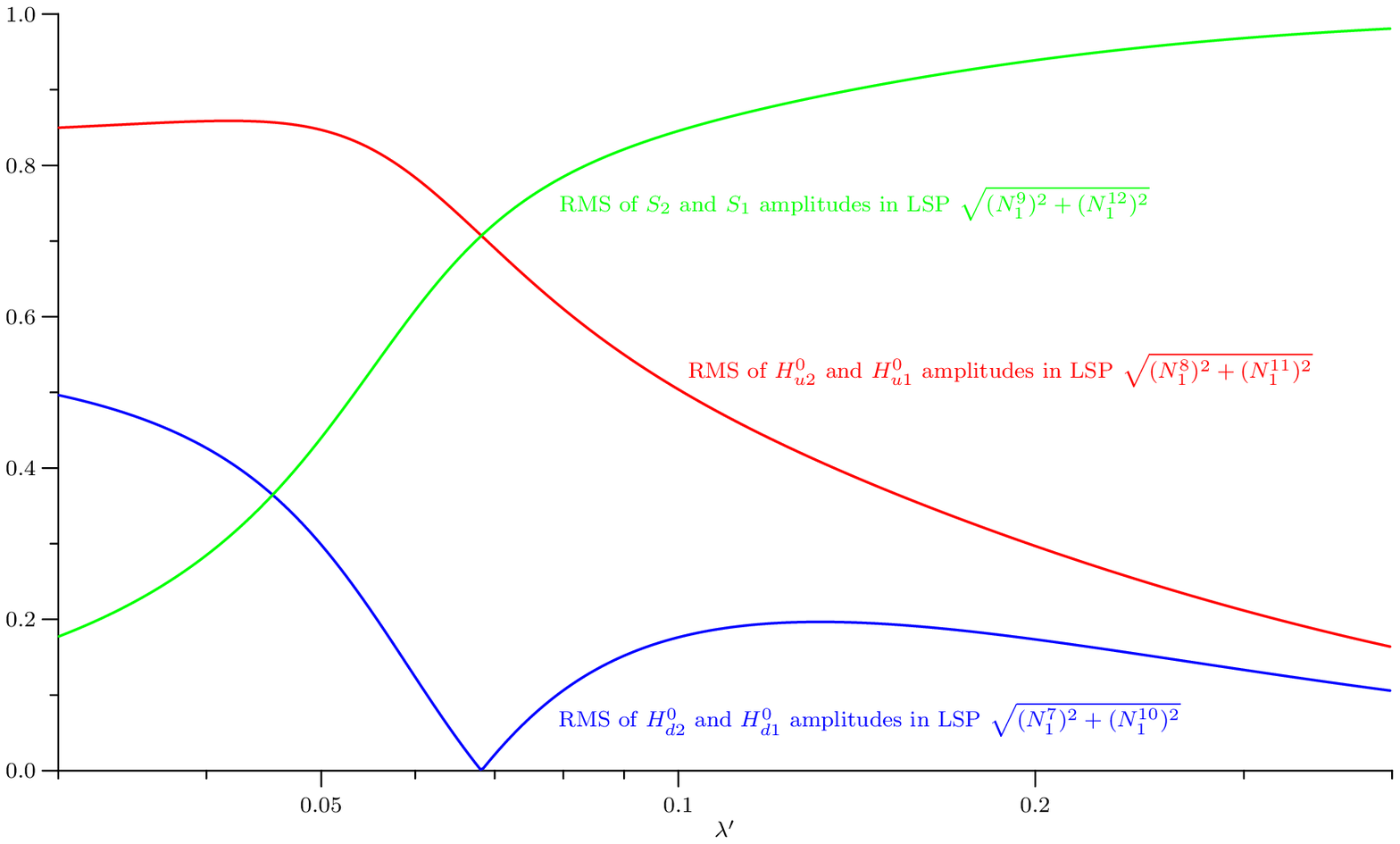}\end{center}
\caption{\it\small Component structure of the LSP
in terms of the inert interaction states
against $\lambda^\prime$
with $f=1$, $\epsilon = 0.1$, $\tan \beta = 1.5$, $s=3000$~GeV and
$Z_2^H$ breaking $\lambda_{ijk}$ couplings set to
0.01.}\label{compLSP}
\end{figure}

Fig.~\ref{specC} shows how the spectrum of chargino masses
varies with $\lambda^\prime$. Although the plot is for $\tan \beta =
1.5$, as one can see from the chargino mass
matrix, Eq.~(\ref{cmm}), the inert sector has no
dependence on $\tan \beta$, with the mass terms just being
proportional to the singlet VEV. The almost constant masses are
those mass eigenvalues coming mostly from the USSM sector, the
third generation charged Higgsino and wino. The charginos coming
mostly from the inert sector vary with $\lambda^\prime$ as
expected and drop below the 94~GeV experimental lower limit at
some value of $\lambda^\prime$, depending on the value of $s$.
The effect of the $\epsilon=0.1$ mixing between generations can bee seen
in the splitting between the two inert sector charginos. Where lines
cross in Fig.~\ref{specC} the chargino masses are of opposite sign.
When chargino mass lines of the same sign approach each other, they veer away from each other
at the would-be crossing point due to the effect of interference.

Fig.~\ref{specN} shows how the spectrum of neutralino
masses varies with $\lambda^\prime$. The inert neutralino spectrum
is dependent on $\tan \beta$, but each of the qualitative
features can be understood. We see the two light neutralino
states that become heavier as $\lambda^\prime$ decreases from
unity until the approximation $\lambda^\prime s \gg fv$ breaks down. At this
point $fv\sin \beta$ begins to dominate and the LSP mass
decreases with decreasing $\lambda^\prime$ as the dominance of
$fv\sin \beta$ becomes greater. In this low $\lambda^\prime$
region the LSP is no longer mostly inert singlino, but mostly
inert up-type Higgsino. The six almost unvarying neutralino masses are
those mostly from the USSM sector, which is not mixing very much
with the new sector in this figure. We have already seen that
the inert sector chargino
masses continue to be set by $\lambda^\prime$ as we go down into
the low $\lambda^\prime$ region, resulting in light charginos in
this region. By contrast, the four inert sector
neutralinos begin to be governed by the $fv$ terms rather than
the $\lambda^\prime s$ terms in the low $\lambda^\prime$ region and
therefore approach a constant value in this region.

As in the case of the charginos,
the effect of the $\epsilon=0.1$ mixing can be seen in the splitting
between the two light neutralinos and the four heavier inert
neutralinos which are both split by this mixing and further split
by the light neutralino mass as predicted in the previous section.

Fig.~\ref{compLSP} shows how the make-up of the LSP in terms of
the inert interaction states varies with $\lambda'$.
The behaviour in the $\lambda' s \gg fv$ limit is as predicted in
Eq.~(\ref{N1}). We also see how the dominant component of the
LSP changes from inert singlino to inert up-type Higgsino
in the low $\lambda'$ region.

\subsection{Dark Matter Relic Density Predictions}

\begin{figure}[t]
\begin{center}
\includegraphics[width=150mm]{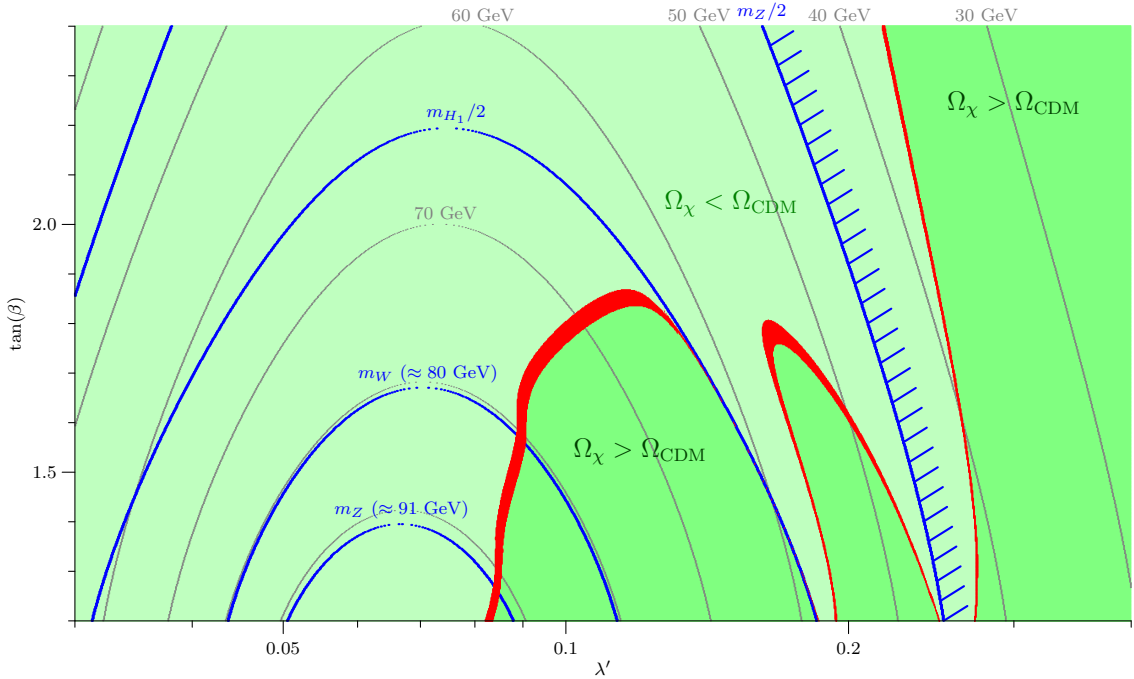}\end{center}
\caption{\it\small Contour plot of the LSP mass
and relic density $\Omega_\chi h^2$ regions in the
$(\lambda^\prime,\tan \beta)$-plane with $s=3000$~GeV, $\epsilon=0.1$ and
$f=1$. The red region is where the prediction for $\Omega_\chi
h^2$ is consistent with the measured 1-sigma range of
$\Omega_\rr{CDM}h^2$. Where the LSP is lower than half of the
$Z$-boson mass, the region to the right of the hatched line
is ruled out by $Z$ decay data.}\label{mBeta}
\end{figure}

\begin{figure}[t]
\begin{center}
\includegraphics[width=150mm]{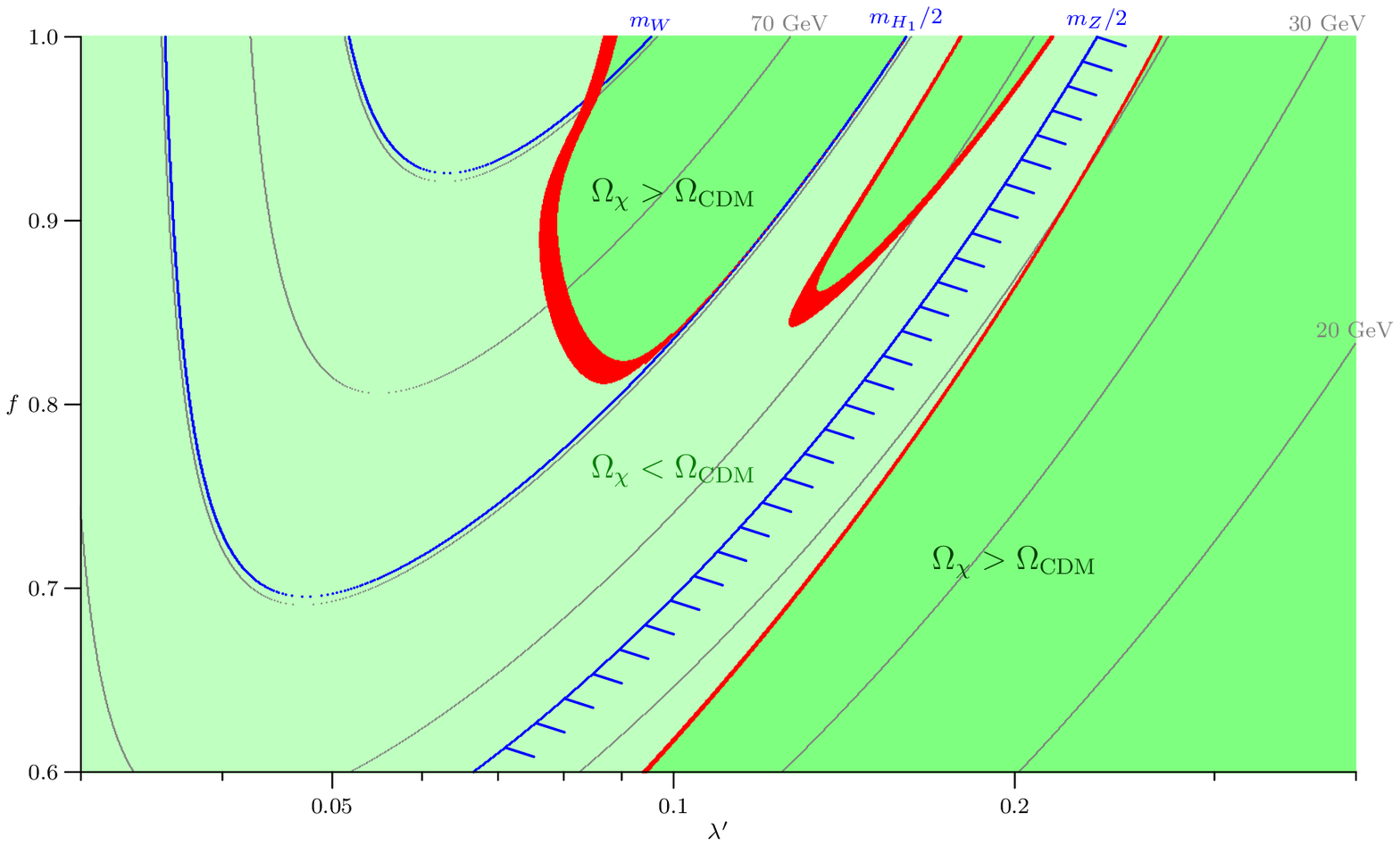}
\end{center}
\caption{\it\small Contour plot of the LSP mass
and relic density $\Omega_\chi h^2$ regions in the
$(\lambda^\prime,f)$-plane with $s=3000$~GeV, $\epsilon=0.1$ and
$\tan \beta=1.5$. The red region is where the prediction for
$\Omega_\chi h^2$ is consistent with the measured 1-sigma range of
$\Omega_\rr{CDM}h^2$. Where the LSP mass is lower than half of the
$Z$-boson mass, the region to the right of the hatched line
is ruled out by $Z$ decay data.}\label{m}
\end{figure}

Using the parametrization in Eqs.~(\ref{param1},\ref{param2}) we
use {\tt MicrOMEGAs 2.2} to numerically compute the
present day relic density.
Fig.~\ref{mBeta} shows a contour plot of the LSP mass
and relic density $\Omega_\chi h^2$ regions in the
$(\lambda^\prime,\tan \beta)$-plane, with $s=3000$~GeV, $\epsilon=0.1$ and
$f=1$. We focus on small values of $\lambda'<0.4$ since for large $\lambda'$
the LSP would be very light state, predominantly inert singlino, which would not annihilate
very efficiently through any channel, leading to a too high relic density
$\Omega_\chi h^2 >\Omega_\rr{CDM}h^2$ (such regions are shaded dark green).
As $\lambda'$ is decreased below 0.3 the LSP mass increases and approaches about half of the $Z$-boson mass
and there is a region where the prediction for $\Omega_\chi
h^2$ is consistent with the measured 1-sigma range of $\Omega_\rr{CDM}h^2$
(such regions are shaded red).
When the LSP mass is around 40~GeV it contains enough inert doublet Higgsino
component such that s-channel $Z$-boson annihilation
becomes strong enough to account for the observed the relic density.
As the LSP mass is increased further from 40~GeV
and approaches 45~GeV, the annihilations before freeze-out
become on resonance with a $Z$-boson in the s-channel and the predicted relic density becomes too low
(such regions are shaded light green).

However the regions where the LSP mass is
less than half of the $Z$-boson mass are excluded by LEP limits on the $Z$-boson width.
The point is that the same couplings which lead to successful relic density,
via annihilation through an s-channel $Z$-boson, will also violate the LEP collider limits
on the $Z$-pole arising from $Z$-boson decay two LSPs. Such a $Z$-boson decay channel
would contribute to the invisible $Z$ width \footnote{By contrast $Z$ decays involving
the second lightest neutralino would contribute to the total width,
because the second neutralino would decay to the LSP
before reaching the detector.}. The measurement of the invisible
$Z$ width at LEP is used to give strong bounds on the number of
light neutrino species \cite{p2}. The PDG average
for the effective number of light neutrinos as inferred from the
invisible $Z$ width is $2.92 \pm 0.05$ \cite{p2}. Because of the
coupling suppression of the LSP due to its inert singlino
component amplitudes, helicity suppression and also
significant phase-space suppression, the branching ratio to two
LSPs would have a contribution to the invisible width significantly less
than that of a neutrino, but still large enough to violate the LEP limit.
Note that in the MSSM this limit does not arise since either the LSP is bino-like
and so does not couple to the $Z$ or is Higgsino or Wino like in which case it would
have accompanying almost degenerate charginos and therefore must have a mass greater
than about 100 GeV in any case. Here we can have an inert Higgsino/singlino LSP with a mass
lower than half of the $Z$-boson mass while still having experimentally
consistent inert-doublet-Higgsino-like charginos.
The regions in Fig.~\ref{mBeta}
where the LSP is lower than half of the
$Z$-boson mass, namely to the right of the hatched line,
are therefore ruled out by the $Z$ decay width measurements at LEP.
Fortunately there are successful regions indicated in red to the left
of the hatched line in Fig.~\ref{mBeta}, where the LSP mass is greater than 45 GeV thereby
avoiding the LEP limit, as we discuss below.

We note at this point that the requirement that the LSP mass exceeds 45 GeV implies
low $\tan \beta$, and this is the reason for the restricted range of $\tan \beta$
in Fig.~\ref{mBeta}. This can be seen from Eq.~(\ref{sin2beta}) where we found that the LSP
mass should be approximately proportional to $\sin 2\beta$,
i.e.~to the product of the two doublet Higgs VEVs, which is maximized for
$\sin 2\beta = 1$ corresponding to $\tan \beta=1$.
In the E$_6$SSM an experimentally acceptable
lightest Higgs mass can be achieved even with $\tan \beta$ as low
as about 1.2 \cite{King:2005jy}, so having low $\tan \beta$ is not a problem in such models.

Decreasing $\lambda'$ further results in LSP masses above 45 GeV, and
to the left of the hatched line in Fig.~\ref{mBeta},
other successful relic density regions (shaded in red) appear.
These regions are punctuated by the light Higgs resonance, leading to
the interesting double loop shape of the successful red regions to the left of the hatched
line in Fig.~\ref{mBeta}.
In these regions the LSP can have a mass significantly larger than half of the $Z$
mass, moving far enough off the Higgs and $Z$ resonances that annihilation is weakened
just enough to give the correct relic density.

However another effect
comes into play as $\lambda'$ decreases, namely the composition
of the LSP changes from being singlino dominated to being Higgsino
dominated, the cross-over point being close to $\lambda'=0.07$,
according to Fig.~\ref{compLSP}.
Within a successful region one would normally expect the relic density to
increase as the LSP mass goes up
(corresponding to decreasing $\tan \beta$ or $\lambda'$) because annihilation moves
further away from the particular resonance (either Higgs or $Z$).
However, for lower $\lambda'$ the cross-section actually increases with decreasing $\lambda'$,
leading to a lower relic density, because the inert doublet Higgsino components in the
LSP rapidly grow, as can be seen in Fig.~\ref{compLSP}.
This implies that for $\lambda'<0.07$, when the LSP is largely
inert doublet Higgsino, annihilation is too strong leading to the relic
density being too low (as indicated by the light green shading in Fig.~\ref{mBeta}.
Note also that here the analytic approximations based on $\lambda^\prime s \gg fv$
break down, leading to the turning over of the LSP mass contours
for $\lambda'<0.08$.
The effects of the t-channel
$W$ and then $Z$ pair production channels can also be seen as they
each become relevant.

According to the above discussion the successful regions to the left of the hatched line in Fig.~\ref{mBeta}
are not ruled out by $Z$ decay data, as the LSP is sufficiently heavy.
Furthermore, for the entire successful region, the lightest chargino is
heavy enough to be consistent with experiment, as can be seen on Fig.~\ref{specC}.
This result will be recreated
for all high enough values of $s$. For larger values of $s$ the
successful regions and corresponding inert chargino masses are shifted down by
the same amount in $\lambda'$.

When $\lambda^\prime s \gg fv$ lowering $f$
results in a lower LSP mass, as in Eq.~(\ref{m1}). It also extends the range of
$\lambda^\prime$ in which this approximation is valid, i.e.~it moves
the boundary of the previously discussed
low $\lambda'$ region to be further down in $\lambda'$.
Fig.~\ref{m} shows the LSP mass and predicted present day relic
density for different values of $\lambda^\prime$ and $f$ with
$\epsilon = 0.1$ and $\tan \beta=1.5$. The shifting of the
successful region, where the
LSP mass is above $m_Z/2$, down in $\lambda'$ at lower values of $f$ is apparent.
At lower values of $\tan \beta$ this successful region
extends further down in $f$. It should be
noted that in order to predict the correct dark matter relic
density, $\lambda^\prime$ should be smaller than $f$ and that this
disparity gets greater if $s$ is increased.
Increasing $s$ effectively just shifts all of the features on
Fig.~\ref{mBeta} and Fig.~\ref{m} to the left.

\subsection{Deviations from the Considered Parametrisation}

Breaking the relation $f_{u(22,11)}=f_{d(22,11)}$ can have similar effects
to those of changing $\tan \beta$. However, because these
parameters cannot be too high (in order to be consistent with
running from the grand unification scale) and because lowering
them to much less than unity makes the LSP too light,
$\tan \beta$ can be varied much more freely than the $f_u/f_d$ ratio.

The effect of increasing the generation mixing parameter $\epsilon$
is to increase the various mass splittings between similar inert
mass eigenstates. Increasing the mixing between the first and
second generations thus results in a lighter LSP, shrinking the
successful region, and
a lighter lightest chargino, potentially inconsistent with
current chargino non-observation.

Increasing the $Z_2^H$ breaking $\lambda_{ijk}$ couplings
from 0.01, it is possible to give
the LSP significant components of the conventional, non-inert
doublet Higgsinos and third generation singlino. However, turning up
these parameters does not allow for the result of a very light
LSP, usually singlino dominated, to be avoided,
simply because of the non-diagonal structure
of the non-gaugino part of the neutralino mass matrix.
Turning up these parameters would, however,
mean that the LSP could have
significant couplings to regular quarks and leptons.

Other parameters only change the neutralinos and charginos mostly
from the USSM sector. As long as the LSP is still mostly from the
inert sector, as considered here (gaugino masses
cannot be too light or else the LSP can become bino/bino$'$
dominated), these parameters are effectively free.
Squark and slepton parameters do not affect the dark matter
physics of the considered model. Top and stop loops can have a
significant effect on the lightest Higgs mass, but as long as this
mass is experimentally allowed then these parameters are also
effectively free.

\section{Summary and Conclusions}
\label{conclusion}

In this paper we have studied
neutralino dark matter arising from supersymmetric models with extra inert Higgsinos
and singlinos. As an example, we have considered the
extended neutralino sector of the E$_6$SSM, which predicts three
families of Higgs doublet pairs, plus three singlets, plus a $Z'$, together
with their fermionic superpartners which include two families of inert Higgsinos and singlinos.
In our study we have considered neutralino dark matter arising from such a model
both analytically and numerically, using {\tt MicrOMEGAs}.

We have found that the results for the relic abundance
in the E$_6$SSM are radically different from those for both the MSSM
and the USSM. This is because the two families of inert Higgsinos
and singlinos predicted by the E$_6$SSM provide an almost
decoupled neutralino sector with a naturally light LSP which can account for
the cold dark matter relic abundance independently of the
rest of the model.
Although the E$_6$SSM has two inert families, the presence of the second
inert family is not crucial for achieving successful dark matter relic abundance.

In the successful regions where the observed dark matter relic density is reproduced
the neutralino mass spectrum is well described by
the analytical results of Section~\ref{analytical}.
In this region the LSP is mostly inert singlino and has a mass
approximately proportional to $v^2/s$, as in Eq~(\ref{m1}),
and, as $\lambda's$ is decreased,
the LSP becomes heavier and also less inert singlino dominated, picking up
significant inert doublet Higgsino contributions.
To avoid conflict with high precision LEP data on the
$Z$-pole, the LSP, which necessarily must couple significantly to the
$Z$-boson in order to achieve a successful relic abundance,
should have a mass which exceeds half the $Z$-boson mass.
Since the LSP mass in Eq.~(\ref{m1}) is proportional to $f_d f_u \sin(2\beta)$,
we find that regions of parameter space in which the dark matter relic density
prediction is consistent with observation require
low values of $\tan \beta$, less than about 2. Depending on the
value of the singlet VEV $s$,
the $f_{u,d}$ trilinear Higgs coupling parameters should also be
reasonably large compared to the
$\lambda_{\alpha\beta}$ ones.
In general it is difficult
for the true neutralino LSP to be heavier than about 100 GeV. In the successful regions
we find the lightest chargino mass could
be as low as the experimental lower limit of 94~GeV, although it could
also be as high as about 300~GeV.

One of the main messages of this paper is that neutralino dark
matter could arise from an almost decoupled sector of inert
Higgsinos and singlinos, and if it does then the parameter space
of the rest of the model is completely opened up. For example if
such a model is regarded as an extension of the MSSM, then
the lightest MSSM-like SUSY particle is not even required to be a
neutralino, and could even be a sfermion which would be able to
decay into the true LSP coming from the almost decoupled inert
Higgsino/singlino sector. This is because the mostly
inert-Higgsino/singlino LSP would have admixtures of MSSM
neutralino states. The size of these components are set by $Z_2^H$
breaking $\lambda_{ijk}$ couplings and need not be extremely
small.

Finally we remark that,
although we have focussed on the E$_6$SSM,
similar results should apply to any singlet-extended SUSY model
with an almost decoupled inert Higgsino sector with a trilinear Higgs coupling
as in Eq.~(\ref{lambda}).

\section*{Acknowledgements}
We are very grateful to Jonathan Roberts for his invaluable help with
the writing of the {\tt LanHEP} code for the model considered in this paper
and with its implementation into {\tt MicrOMEGAs}.
The {\tt LanHEP} code for the considered E$_6$SSM scenario is an extension of his
code for the complete USSM, which was used for Ref.~\cite{Kalinowski:2008iq},
and we thank Jonathan Roberts and Jan Kalinowski for donating this code, and
for critically reading this manuscript.
We would also like to thank A.~Belyaev for fruitful
discussions. SFK acknowledges partial support from the following
grants: STFC Rolling Grant ST/G000557/1; EU Network
MRTN-CT-2004-503369; NATO grant PST.CLG.980066; EU ILIAS
RII3-CT-2004-506222.

\end{document}